          [2001/04/25 1.1 (PWD)]
\documentclass[a4paper,twocolumn]{esapub} 
\usepackage{times}
\usepackage{natbib}
\usepackage{graphicx}
\usepackage{amsmath}

\def\HeII{He\,{\scriptsize II}}

\title{On seismic signatures of rapid variation}
\author[1]{G. Houdek}
\affil[1]{Institute of Astronomy, University of Cambridge, 
          Cambridge CB3 0HA, UK}
\author[1,2]{D. O. Gough}
\affil[2]{Department of Applied Mathematics and Theoretical Physics, 
       University of Cambridge, Cambridge CB3 0WA, UK}

\begin{document}


\maketitle

\begin{abstract}
We present an improved model for an asteroseismic diagnostic contained in the
frequency spacing of low-degree acoustic modes. By modelling in a realistic 
manner regions of rapid 
variation of dynamically relevant quantities, which we 
call acoustic glitches, we can derive signatures of the gross properties of 
those glitches.  In 
particular, we are interested in measuring properties that are related to the 
helium ionization zones and to the rapid variation in the background state 
associated with the lower boundary of the convective envelope. The 
formula for the seismic diagnostic is tested against a sequence of 
theoretical models of the Sun, and is compared with seismic 
diagnostics published  previously by Monteiro \& Thompson (1998, 2005) 
and by Basu\,et\,al. (2004).

\end{abstract}

\section{Introduction}

Abrupt variation in the stratification of a star (relative to the scale of the 
inverse radial wavenumber of a seismic mode of oscillation), such as that 
resulting from the (smooth, albeit acoustically relatively abrupt) depression 
in the first adiabatic exponent 
$\gamma_1 = (\partial {\ln p}/\partial{\ln\rho})_s$ 
caused by the ionization of helium, 
where $p$, $\rho$ and $s$ are pressure, density and specific entropy, 
or from the 
sharp transition from radiative to convective heat transport at the base of 
the convection zone, induces small-amplitude oscillatory components (with respect to 
frequency) in the spacing of the cyclic eigenfrequencies $\nu_{n,l}$ of seismic oscillation. 
One might hope that the variation of the sound speed induced by helium 
ionization might enable one to determine from the low-degree 
eigenfrequencies a measure that is directly related to, perhaps even almost 
proportional to, the helium abundance, with little contamination from other 
properties of the structure of the star.

A convenient and easily evaluated measure of the oscillatory component is the 
second multiplet-frequency difference with respect to order $n$ amongst modes 
of like degree $l$:

\begin{equation}
\Delta_2\nu_{n,l}\equiv\nu_{n-1,l}-2\nu_{n,l}+\nu_{n+1,l}\,
\end{equation}

(Gough, 1990). Any 
localized region of rapid variation of either the sound speed or the density 
scale height, or a spatial derivative of them, which here we call an acoustic 
glitch, induces an oscillatory component in  $\Delta_2\nu$ with a 
`cyclic frequency' approximately equal to twice the acoustic depth

\begin{equation}
\tau=\int_{r_{\rm glitch}}^R c^{-1}\,{\rm d}r
\end{equation}

of the glitch, and with an amplitude which depends on the amplitude of the 
glitch and which decays with $\nu$ once the inverse radial wavenumber of the 
mode becomes comparable with or greater than the radial extent of the glitch. 

We report here on a model formula for the oscillatory contribution to the 
second frequency difference that is directly related to the gross properties 
of the glitch, and compare how well it fits actual eigenfrequency differences 
with corresponding fits of previously published formulae.

\section{The fitting function}

For a nonrotating star with vanishing pressure gradient at the surface, the 
angular eigenfrequencies $\omega$ of adiabatic oscillation obey a variational 
principle 
(e.g. Chandrasekhar 1963) which can be used to estimate the contribution 
$\delta\omega$ to $\omega$ from the acoustic glitch. In estimating the 
contribution from helium ionization we retain only the dominant glitch 
term ($\delta\gamma_1/\gamma_1$ at fixed pressure and temperature), and we 
represent $\delta\gamma_1/\gamma_1$ by a pair of (negative) Gaussian 
functions of $\tau$ with widths $\Delta_{\rm I}$ and $\Delta_{\rm II}$, whose 
integrals are $\Gamma_{\rm I}$ and $\Gamma_{\rm II}$, and which are centred 
about the acoustic 
depths $\tau_{\rm I}$ and $\tau_{\rm II}$  of the first and second ionization 
zones of helium beneath the seismic surface $r=R$ of the 
star. After converting to cyclic frequency $\nu=\omega/2\pi$, the oscillatory 
component is given asymptotically by

\begin{eqnarray}
\delta_{\gamma,{\rm osc}}\nu&\simeq&
    -\Gamma_{\rm II}\nu_0\left[\nu+\tfrac{1}{2}(m+1)\nu_0\right]\nonumber\\
& &\times\bigl(\mu\beta\kappa_{\rm I}^{-1}
 {\rm e}^{-8\pi^2\mu^2\kappa^2_{\rm I}\Delta_{\rm II}^2\nu^2}\cos2\psi_{\rm I}\nonumber\\
& &{}\;\;\;+\kappa_{\rm II}^{-1}{\rm e}^{-8\pi^2\kappa^2_{\rm II}\Delta_{\rm II}^2\nu^2}\cos2\psi_{\rm II}\bigr)\,
\end{eqnarray}
(Houdek \& Gough, 2007).  
We evaluate the phases $\psi_{\rm I}=\psi(\tilde\tau_{\rm I})$ and 
$\psi_{\rm II}=\psi(\tilde\tau_{\rm II})$, where 
$\omega\tilde\tau=\omega\tau+\epsilon_{\rm II}$, 
by representing the envelope by a 
plane-parallel polytrope of index $m=$3.5 and adding a phase constant 
$\epsilon_{\rm II}$ to $\omega\tau$ to account for the deviation of the actual 
envelope from the polytrope: 
 \begin{eqnarray}
\psi(\tau)&\!\!\!=\!\!\!&\omega{\tau}{\kappa}-(m+1)\cos^{-1}
\left(\frac{m+1}{{\tau}\omega}\right)+\frac{\pi}{4}~.\,\,
\end{eqnarray}
In equation~(3),  
$\kappa_{\rm I}=\kappa(\tilde\tau_{\rm I})$ etc, with
$\kappa(\tau)=[1-(m+1)^2/4 \pi^2 \nu^2 \tau^2]^{1/2}$; also $\nu_0$ is a 
characteristic value of the first frequency difference $\Delta_1\nu$ whose 
value is half the inverse total acoustic radius $T$ of the star: $\nu_0=
\omega_0/2\pi=1/2T$.
We have found that the ratios
$\beta=\Gamma_{\rm I}\Delta_{\rm II}/\Gamma_{\rm II}\Delta_{\rm I}$, 
$\eta=\tau_{\rm I}/\tau_{\rm II}$ and
$\mu=\Delta_{\rm I}/\Delta_{\rm II}$ hardly vary amongst stellar models whose 
masses and radii vary by factors of at least five. We have accordingly  adopted 
constant values, namely  $\beta=0.45$, $\eta=0.7$ and $\mu=0.9$ for solar-like 
stars.  

To obtain a complete description of $\Delta_2\nu$ we must add an oscillatory  
contribution 
from the near discontinuity in the density scale height at the base 
$\tau_{\rm c}$ of the convection zone:

\begin{eqnarray}
\delta_{\rm c, osc}\nu&\simeq&\hat A_{\rm c}\nu_0^3\nu^{-2}
   \left(1+1/{16\pi^2\tau_0^2\nu^2}\right)^{-1/2}\cr
&\times&\left\{\cos[2\psi_{\rm c}+\tan^{-1}(4\pi\tau_0\nu)]
\right\},
\end{eqnarray}
with
\begin{equation}
{\hat A_{\rm c}}=\frac{c^2_{\rm c}}{8\pi\omega^2_0}
\left[{{\rm d}^2\ln\rho}/{{\rm d}r^2} \right]
^{r_{\rm c}{+}}_{r_{\rm c}{-}}\,,
\end{equation}

where $\psi_{\rm c} =\psi(\tau_{\rm c}+\omega^{-1}\epsilon_{\rm c})$ is the 
polytropic phase at the base of the convection zone, the phase constant
$\epsilon_{\rm c}$ being allowed to differ from $\epsilon_{\rm II}$ to account for
the fact that the region of the convection zone beneath the helium ionization 
zones is not an exact polytrope; the quantity  
$\tau_0$ is a characteristic scale over which the sound 
speed in the radiative interior relaxes to a smooth extrapolation from the 
convection zone (and which we also take to be constant, 80\,s, for all the 
stars we consider).

From these expressions it is straightforward to evaluate the second 
difference $\Delta_2\nu$, to which must be added a smooth term which we 
represent by a third-degree polynomial in $\nu^{-1}$:

\begin{equation}
\Delta_{2,{\rm sm}} = \sum^3_{i=0} a_i \nu^{-i}\,.
\end{equation}

The eleven parameters
$
{A_{\rm II}}, {A_{\rm c}}, {\Delta_{\rm II}}, 
{\tau_{\rm II}}, {\tau_{\rm c}}, \epsilon_{\rm II}, 
\epsilon_{\rm c}\ {\rm and}\ a_i 
$
are adjusted to fit by least squares the theoretical curve to the second 
frequency differences of the actual eigenfrequencies of the modes.

We compare the current seismic diagnostic~(3)--(7) with diagnostics published 
previously by Monteiro \& Thompson (1998, 2005) and by Basu et\,al. (2004).

\subsection{The seismic diagnostic by Monteiro \& Thompson}

This diagnostic is based on a single-triangular approximation to
$\delta\gamma_1/\gamma_1$ due to He ionization.  The resulting formula for 
the frequency perturbation is  
\begin{eqnarray}
\delta_{\rm osc}\nu&\simeq&b_1\frac{\sin^2(2\pi\nu\hat\beta)}{\nu\hat\beta}
\;\cos(4\pi\nu\tau_{\rm II}+2\epsilon_{\rm II})\cr
&+&\left[\frac{c^2_1}{\nu^4}+\frac{c^2_2}{\nu^2}\right]^{1/2}
\;\cos(4\pi\nu\tau_{\rm c}+2\epsilon_{c})\,,
\end{eqnarray}
in which $\hat\beta$ is the half-width  and $b_1$ is 
proportional to the height of the triangle, from which second differences can 
be computed.  
We set to zero the fitting parameter $c_2$, a nonzero value of which, in 
the original 
formulation, would account for an putative adiabatic extension to the convection 
zone produced by overshooting (which is not present in our solar models). 
By so doing, the seismic diagnostic (8) has the 
same number, 11, of fitting parameters as the seismic diagnostic~(3)--(7) 
when adopting the third-degree polynomial (7) to represent the smooth 
contribution.

\subsection{The seismic diagnostic by Basu et\,al.}

In this formulation the seismic diagnostic is that used by Basu (1997), which 
was invented by Basu, Antia and Narasimha (1994) for studying the base of the 
convection zone.  Ballot et al. (2004) and Piau et al. (2005) have adopted 
essentially the same formula.  It is not directly linked to the form
of the ionization-induced acoustic glitch, but instead was motivated by the 
oscillatory signature of a discontinuity in sound speed, whose amplitude was 
modified to make it decay with frequency approximately in the manner of the 
second differences of the eigenfrequencies.  The functional form is given by 
\begin{eqnarray}
\Delta_2\delta_{\rm osc}\nu&\simeq&\left(b_1+\frac{b_2}{\nu^2}\right)
\cos(4\pi\nu\tau_{\rm II}+2\epsilon_{\rm II})\cr
&+&\,\left(c_1+\frac{c_2}{\nu^2}\right)
\cos(4\pi\nu\tau_{\rm c}+2\epsilon_{\rm c})\,.
\end{eqnarray}
Here we set $c_1=0$, to remove the effect of the erroneous sound-speed 
discontinuity at the base of the convection zone and to render the functional 
form of the contribution at $\tau=\tau_{\rm c}$ (nearly) consistent with 
those in the other two 
signatures (5) and (8); and we adopt the third-degree polynomial (7) to 
represent the smooth contribution, thereby using the same number (11) of 
fitting parameters as the seismic diagnostics~(3)--(7) and (8), (7). 
\vspace{-3mm}
\section{Test Models}
\vspace{-3mm}
Two sets of calibrated evolutionary models of the Sun, computed by Gough and 
Novotny (1990), were used. One set of 
models has a constant value for the heavy-element abundance $Z$ but with 
age varying in uniform steps of approximately 0.051 in $\ln t$, and the 
other set has constant age but varying heavy-element abundance in uniform steps
of approximately 0.016 in $\ln Z$ (see Fig.~1).  
\begin{figure}
\centering
\includegraphics[width=0.9\linewidth]{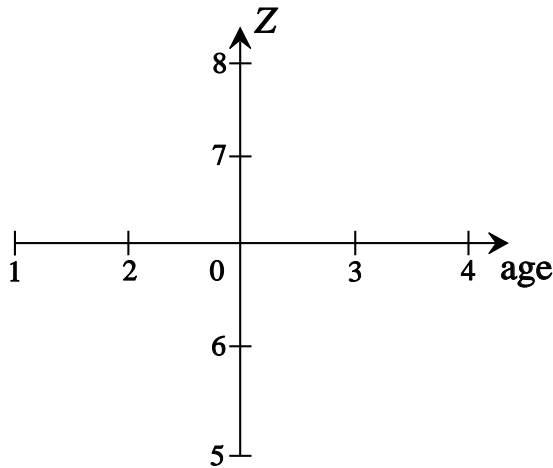}
\caption{Denotation of the nine calibrated solar models which are used for 
testing the formulation of the second frequency differences. The 
`central model' is model 0; the set of the four models (1)--(4) have a constant 
value of $Z$ but varying age; the second set of the models, (5)--(8), have constant 
age but varying $Z$. }
\end{figure}
\vspace{-3mm}
\section{Results}
\vspace{-3mm}
All three diagnostics were fitted by the same numerical algorithm. The lower 
 panels of Figs~2--4 show the results for the three seismic 
diagnostics~(3)--(7), (8) and (7), and (9) and (7), fitted by least-squares to 
the adiabatically computed eigenfrequencies of the nine test models. 
Plotted are the frequency-averaged amplitudes, $\langle A_{\rm II}\rangle$ and 
$\langle A_{\rm c}\rangle$, of the two oscillatory seismic contributions 
associated with \HeII\ ionization and with the base of the convection zone, 
obtained by averaging the appropriate calibrated 
formula uniformly over the frequency range of the data plotted in the upper 
panels.   Also plotted are the 
values for the acoustic depths $\tau_{\rm II}$ and  $\tau_{\rm c}$. Depending
on the seismic diagnostic, the values of either $\Delta_{\rm II}$, $\hat\beta$ or 
the phase constant $\epsilon_{\rm II}$  are shown, and also the standard 
measure $\chi^2$ of the goodness of the fit between $\Delta_2\nu$ of the 
modelled frequencies and the calibrated seismic diagnostic. In all three cases 
the frequency-averaged amplitude $\langle A_{\rm II}\rangle$ increases 
with the helium abundance $Y$ of the test models; moreover, the amplitudes 
$\Gamma_{\rm {II}}$ and  $b_1$ inferred from diagnostics~(3)--(7) and (8), (7)
respectively also increase with $Y$  ($\Gamma_{\rm {II}}$ varying almost 
linearly in the $Z$-varying sequence, $b_1$ varying almost linearly in the 
$t$-varying sequence), the latter result being apparently  
at variance with the findings of Monteiro and Thompson (2005).  The seismic 
diagnostic~(3)--(7)  fits $\Delta_2\nu$ of the model 
eigenfrequencies the best, as is evinced by its smallest values of $\chi^2$. 

The quality of the fits is also illustrated in the upper panels of 
Figs~2--4, which show the second frequency differences over the whole 
frequency range considered for the central model 0. It is also clear from comparing 
these figures that the seismic diagnostic~(3)--(7) reproduces $\Delta_2\nu$ of 
the model frequencies the most faithfully. 

It is interesting to note that in determining the parameters in the formulae 
the fitting procedure gives greater weight to the low-frequency end of the 
ionization-induced component $\Delta_{2,\gamma}$, where the amplitude is the 
greatest, and greater weight to the high-frequency end of the contribution 
$\Delta_{2,\rm{c}}$ from the base of the convection zone, where misfits cannot 
be offset by adjustments to $\Delta_{2,\gamma}$; therefore, the diagnostics 
(8), (7) and (9), (7) fail principally in the mid-range. Some large-scale 
discrepancy can be seen also at high frequency for formula (9), (7)
because $\Delta_{2,\gamma}$ decays with frequency too slowly. We acknowledge 
that these discrepancies might be either smaller or hardly greater than the 
random errors in imminent asteroseismic data; nevertheless, fitting an 
erroneous formula can lead to the inferences from that fitting being biased, 
and should therefore be avoided.

\begin{figure*}
\centering
\includegraphics[width=0.8\linewidth]{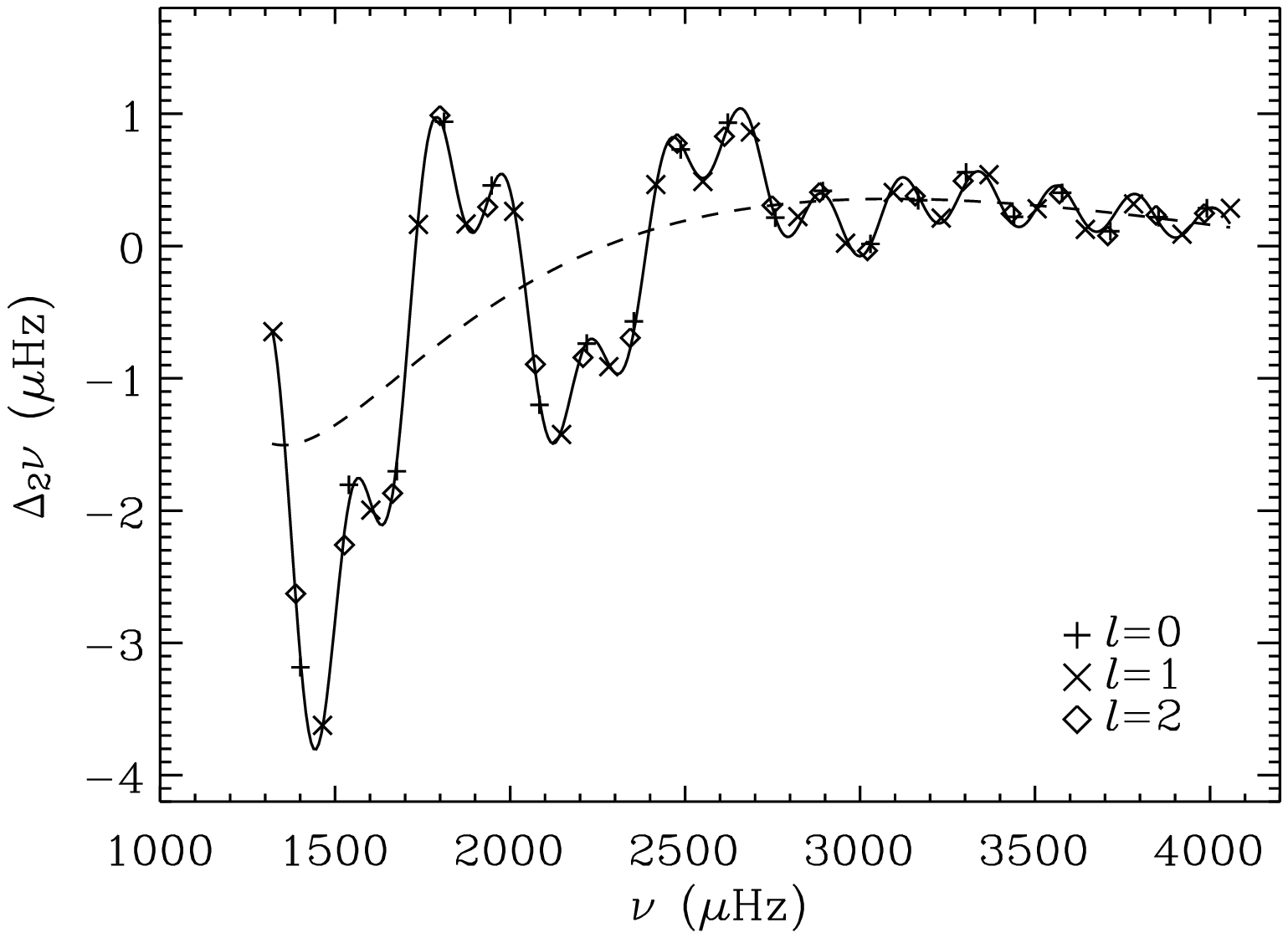}
\includegraphics[width=0.9\linewidth]{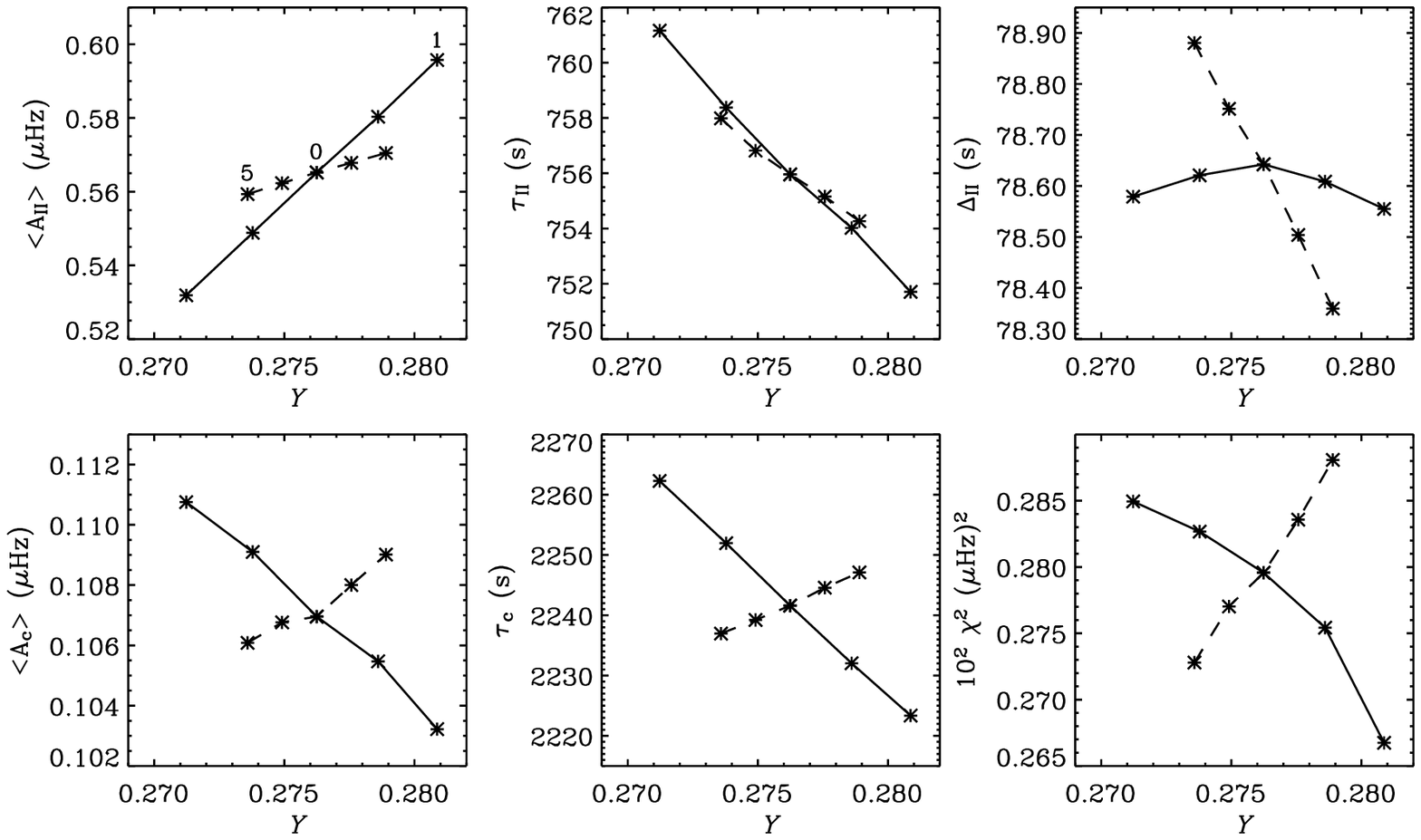}
\caption{ Top: The symbols are second differences $\Delta_2\nu$, 
defined by equation (1), of low-degree frequencies obtained from adiabatic 
pulsation calculations of the central model~0. The curve 
is the seismic diagnostic~(3)--(7) whose eleven parameters have been 
adjusted to fit the data by least squares. Bottom: Model properties 
determined from the seismic diagnostic~(3)--(7) for the nine test models. 
Models with varying age are connected by solid lines, those with varying 
$Z$ by dashed lines.  Some model identifiers are written in the upper left 
panel. 
The smallest (central) frequency value used in the least-square fitting is
$\nu\simeq1322\,\mu$Hz, the largest is $\nu\simeq4058\,\mu$Hz. The lower 
right panel is the standard measure $\chi^2$ for the goodness of the 
least-squares fit. 
} 
\end{figure*} 

\begin{figure*}
\centering
\includegraphics[width=0.8\linewidth]{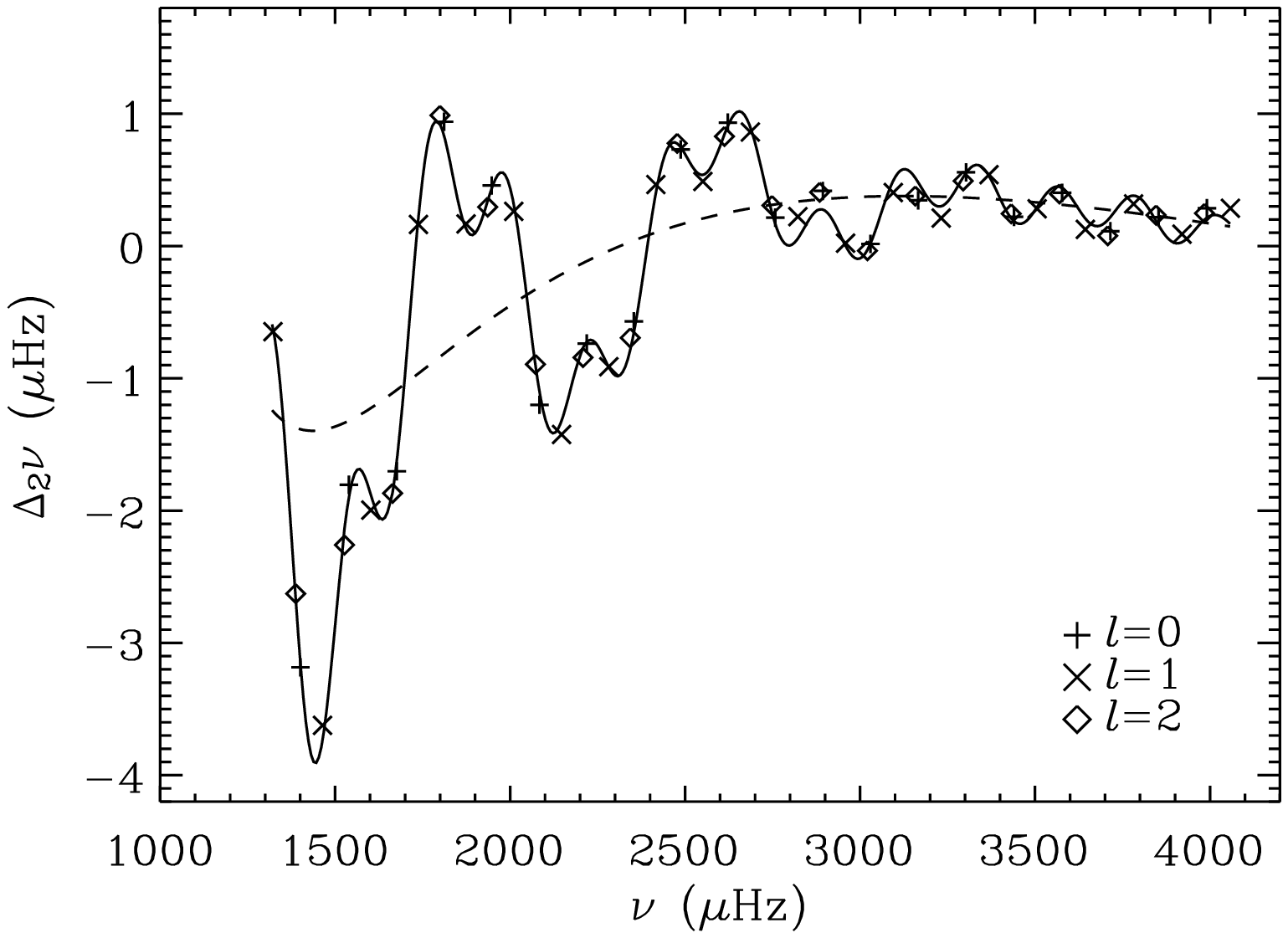}
\includegraphics[width=0.9\linewidth]{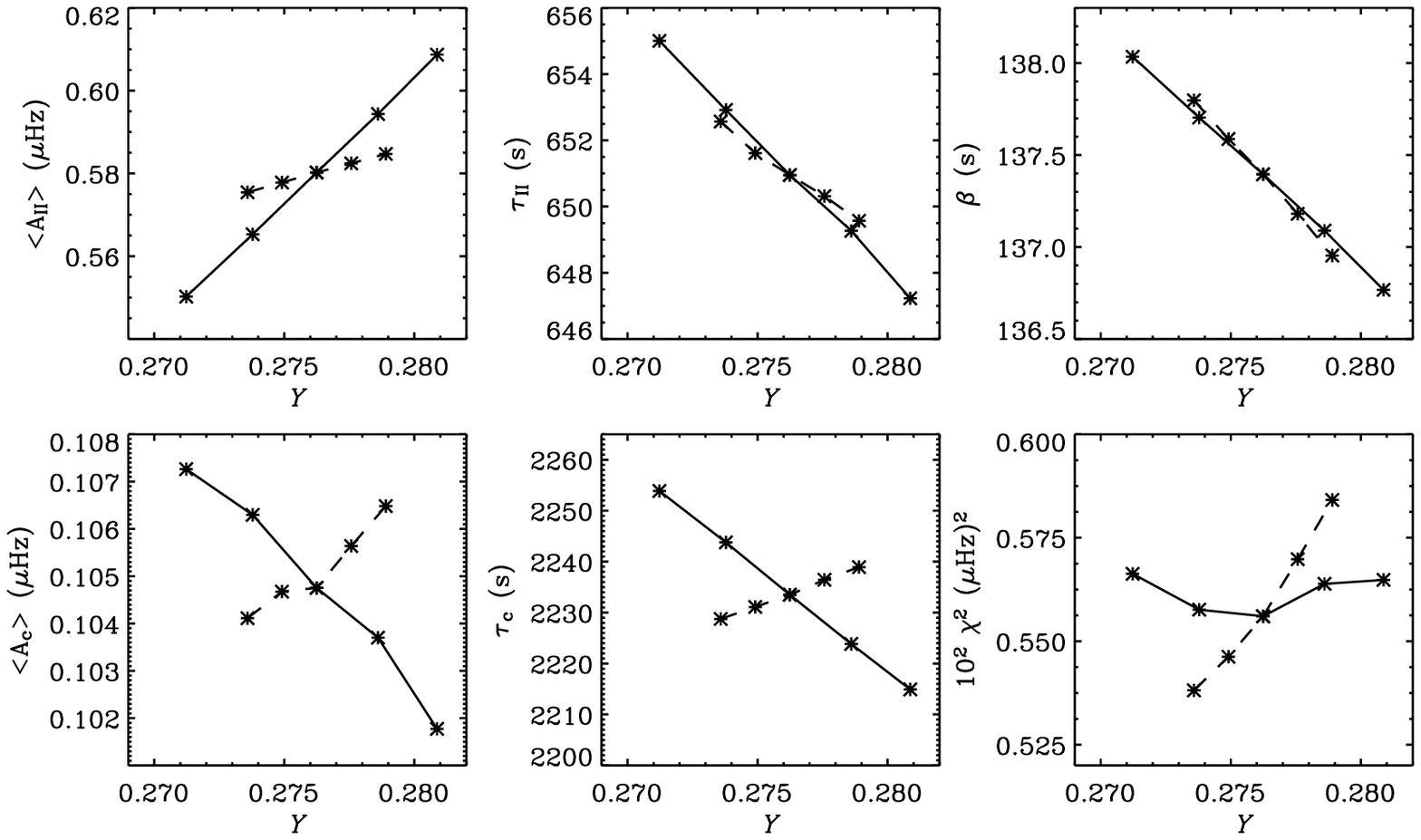}
\caption{ Top: The symbols are second differences $\Delta_2\nu$, 
defined by equation (1), of low-degree frequencies obtained from adiabatic 
pulsation calculations of the central model~0. The curve 
is the seismic diagnostic (8),(7) whose eleven parameters have been 
adjusted to fit the data by least squares. Bottom: Model properties 
determined from the seismic diagnostic (8),(7) for the nine test models. The 
smallest (central) frequency value used in the least-square fitting is
$\nu\simeq1322\,\mu$Hz, the largest is $\nu\simeq4058\,\mu$Hz. The lower 
right panel is the standard measure $\chi^2$ for the goodness of the 
least-squares fit.} 
\end{figure*} 

\begin{figure*}
\centering
\includegraphics[width=0.8\linewidth]{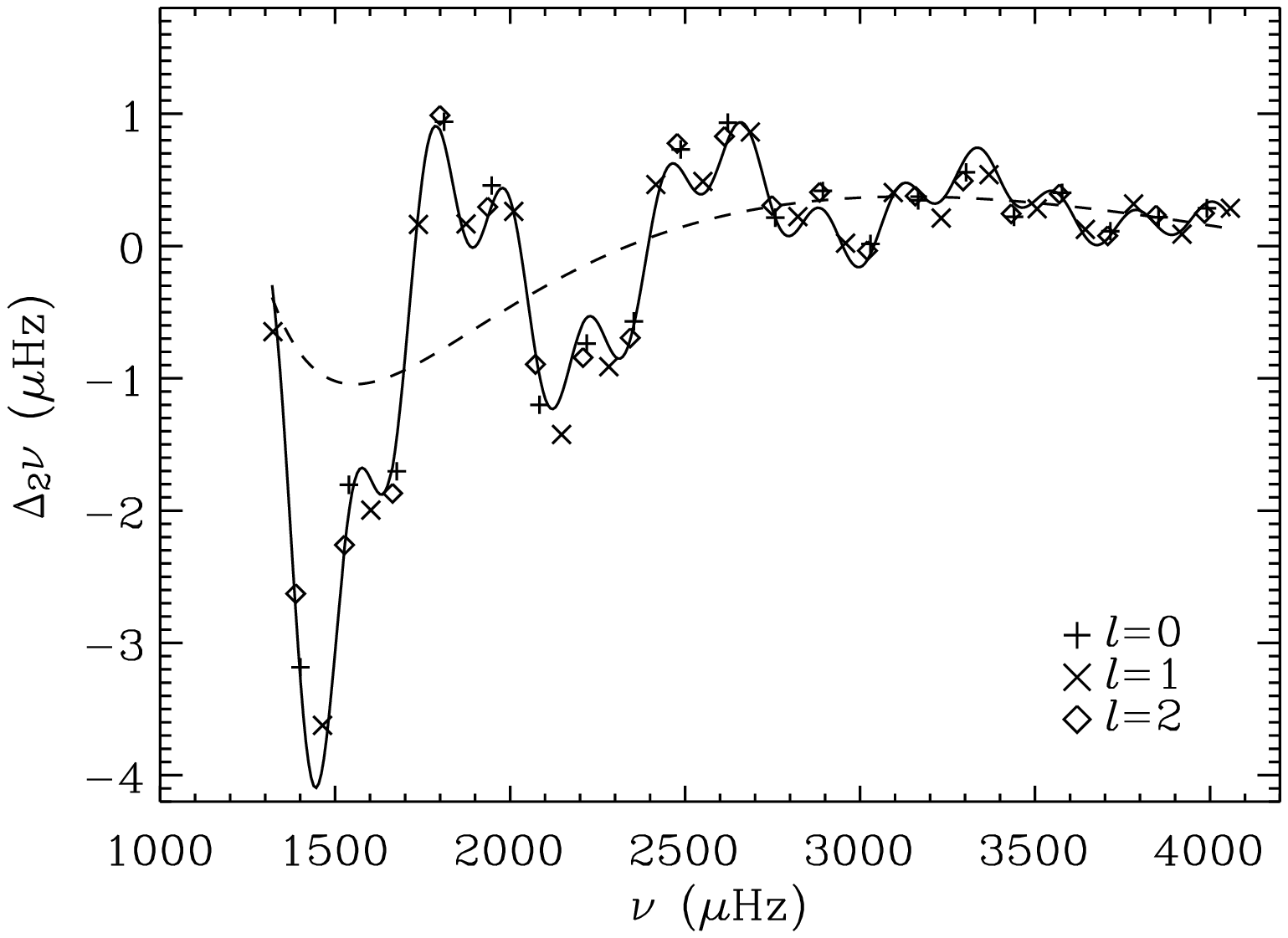}
\includegraphics[width=0.9\linewidth]{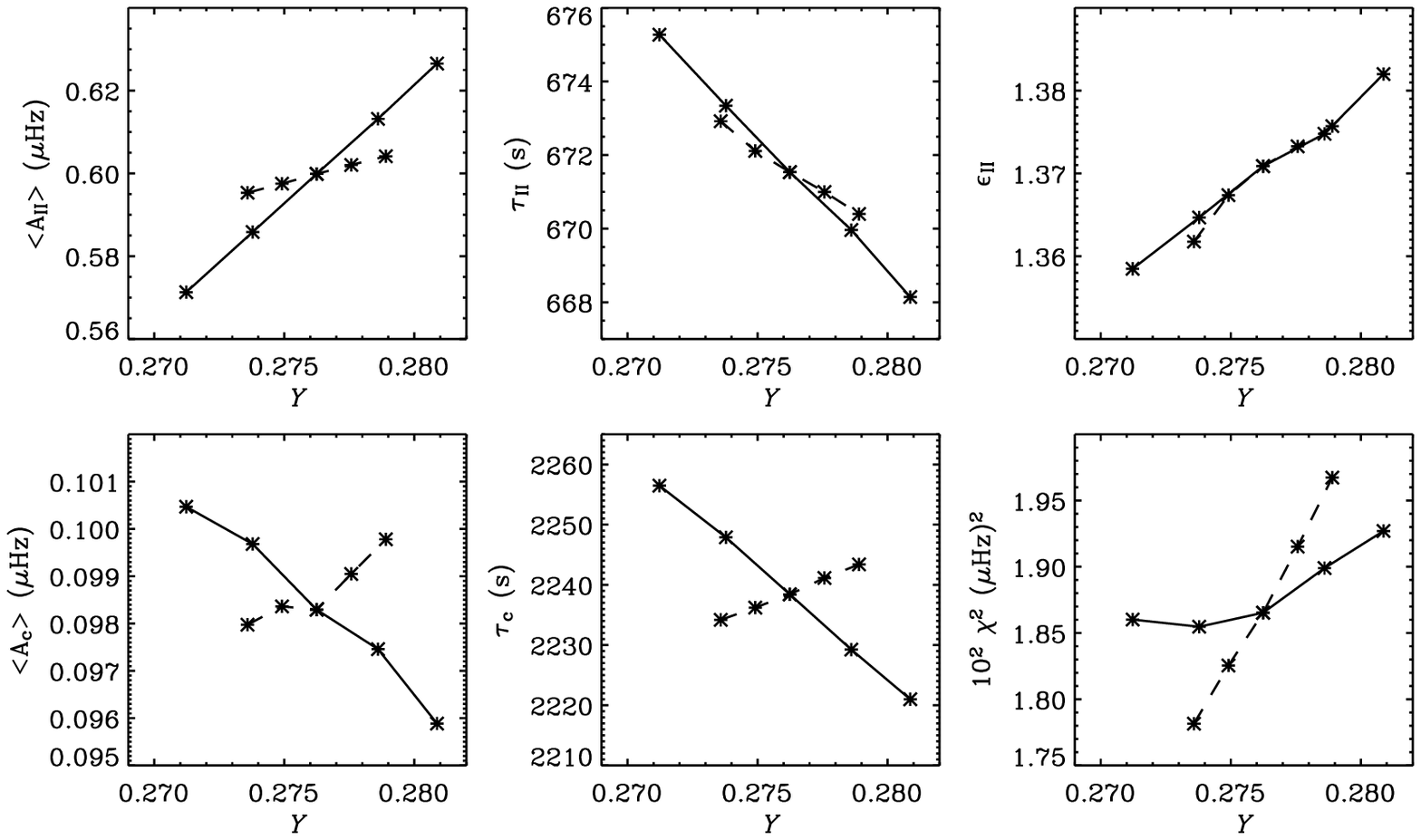}
\caption{ Top: The symbols are second differences $\Delta_2\nu$, 
defined by equation (1), of low-degree frequencies obtained from adiabatic 
pulsation calculations of the central model~0. The curve 
is the seismic diagnostic (9),(7) whose eleven parameters have been 
adjusted to fit the data by least squares. Bottom: Model properties 
determined from the seismic diagnostic (9),(7) for the nine test models. The 
smallest (central) frequency value used in the least-square fitting is
$\nu\simeq1322\,\mu$Hz, the largest is $\nu\simeq4058\,\mu$Hz. The lower 
right panel is the standard measure $\chi^2$ for the goodness of the 
least-squares fit.} 
\end{figure*} 
\vspace{-3mm}
\section*{Acknowledgements}
\vspace{-3mm}
GH is grateful for support from the Particle Physics and Astronomy Research Council.


%
%
%
%
\end{document}